\pdfoutput = 1
\documentclass{aastex62}
\usepackage[utf8]{inputenc}

\shorttitle{Substructure in the M86/M84 GC System}
\shortauthors{Lambert et al.}

\begin{document}

\title{Substructure in the Globular Cluster Populations of the Virgo Cluster Elliptical Galaxies M84 and M86}

\author{Ryan A. Lambert}
\affil{Department of Astronomy, Indiana University\\
Bloomington, IN 47405-7105, USA}

\author{Katherine L. Rhode}
\affil{Department of Astronomy, Indiana University\\
Bloomington, IN 47405-7105, USA}

\author{Enrico Vesperini}
\affil{Department of Astronomy, Indiana University\\
Bloomington, IN 47405-7105, USA}

\nocollaboration

\begin{abstract}
We have carried out a search for substructure within the globular cluster systems of M84 (NGC 4374) and M86 (NGC 4406), two giant elliptical galaxies in the Virgo Cluster. We use wide-field (36$\arcmin$ x 36$\arcmin$), multi-color broadband imaging to identify globular cluster candidates in these two galaxies as well as several other nearby lower-mass galaxies.  Our analysis of the spatial locations of the globular cluster candidates reveals several substructures, including: a peak in the projected number density of globular clusters in M86 that is offset from the system center and may be at least partly due to the presence of the dwarf elliptical galaxy NGC~4406B; a bridge that connects the M84 and M86 globular cluster systems; and a boxy iso-density contour along the southeast side of the M86 globular cluster system. We divide our sample into red (metal-rich) and blue (metal-poor) globular cluster candidates to look for differences in the spatial distributions of the two populations and find that the blue cluster candidates are the dominant population in each of the substructures we identify. We also incorporate the measurements from two radial velocity surveys of the globular clusters in the region and find that the bridge substructure is populated by globular clusters with a mix of velocities that are consistent with either M86 and M84, possibly providing further evidence for interaction signatures between the two galaxies.

\end{abstract}

\section{Introduction} \label{sec:intro}
According to the current paradigm, massive galaxies formed hierarchically, through the continuous merging and accretion of smaller proto-galaxies.  In this picture, massive structures like giant galaxies started to form in the early Universe when small fluctuations in the matter density distribution began to collapse and accrete surrounding matter. These structures continued to merge and build up larger structures over the course of cosmic time.  Simulations that involve this type of hierarchical merging and matter accretion within the context of a $\Lambda$CDM universe have been fairly successful at recreating the observed overall distribution of matter \citep[e.g.,][]{springel05,vogelsberger14,schaye15}
and producing galaxies with realistic properties that are generally well-matched to the properties observed today \citep[e.g.,][]{springel08,hopkins14,hopkins18}. 

Much of the recent work on this topic has focused on gathering observational evidence for hierarchical growth through the identification of stellar streams and tidal interactions, which are thought to be the hallmark of low-mass galaxies having been accreted by more massive galaxies.  The Sagittarius dwarf galaxy \citep{ibata94} and the stellar streams associated with it \citep{belokurov06} are some of the most well-studied examples of this type of hierarchical accretion in the Milky Way.  Recent data from the Gaia mission  \citep{gaia16} and other large-scale surveys, and analyses that combine kinematic data with information about chemical abundances and/or specific stellar populations, have led to discoveries of additional substructures in the Galaxy and brought renewed attention to this topic \citep[e.g.,][]{belokurov18,deason18,myeong18,helmi18}. 
The Andromeda galaxy (M31) has been studied in detail as well. Structures formed from spatial over-densities of the Red Giant Branch (RGB) population, e.g. the Giant Stellar Stream, have been found in the outer halo of M31 by using resolved RGB star maps \citep{ibata01,ferguson02}. Using the surface density of resolved RGB stars, it is possible to infer the surface brightness distribution of the underlying light; structures revealed by this technique, such as faint streams and clumps of RGB stars, suggest a rich history of accretion events and tidal interactions for M31. Deep observations of galaxies outside the Local Group have also revealed streams, shells, and satellites using resolved RGB stars as well as unresolved low surface brightness (LSB) features \citep[e.g.,][]{janowiecki10,crnojevic16,mihos17}. However, in galaxies well beyond the Local Group, RGB stars are unresolved and a different tracer of hierarchical assembly processes and tidal interactions is required. Globular clusters (GCs) have a number of properties that make them well-suited for this task.  

GCs are compact stellar systems with typical masses ranging from $\sim$10$^{4}$ M${_\odot}$ to $\sim$10$^{6}$ M${_\odot}$. They are highly luminous objects and therefore visible up to distances of hundreds of Mpc \citep[e.g.,][]{blakeslee99,mieske04}.  They are ubiquitous across all galaxy types, with giant galaxies typically hosting hundreds or thousands of GCs \citep[e.g.,][]{rhode12,young16}. Furthermore, GC metallicities have been shown to contain information about hierarchical galaxy assembly processes.  The Milky Way has two populations of GCs with different mean metallicities, and the abundances and other properties of these two populations have been used to probe the Galaxy's past history \citep{zinn85,az88}. Most other massive galaxies have GC systems that show multiple peaks in their color distributions \citep[e.g.,][and references therein]{brodie06}.  For integrated colors of stellar populations older than $\sim$2$-$6~Gyr, these color differences can indicate a difference in metallicity, with bluer colors corresponding to lower metallicities and redder colors corresponding to higher metallicities \citep[e.g.,][]{young16}.  The presence of multiple color peaks in the GC systems of some galaxies is commonly interpreted as evidence that such galaxies have undergone two or more major epochs of star formation, possibly triggered by mergers or accretion events \citep[e.g.,][]{az92,za93,li14,elbadry19}. Moreover, the spatial distributions of the GC subpopulations have been shown to differ in many galaxies, with blue/metal-poor GCs typically covering a larger radial extent than their red/metal-rich counterparts \citep[e.g.,][and references therein]{brodie06}. While this spatial difference may have multiple causes, one hypothesis is that a large proportion of the metal-poor GCs in massive galaxies were deposited through accretion events \citep{cote98,forbes10}. 

GCs have also been shown to trace substructure within and around galaxies. For example, \citet{lim17} performed a study of the GCs in the early-type "shell" galaxy NGC~474 and found a significant correlation between the shell and stream structures and the GC spatial positions. Furthermore, previous studies have reported similar results for the M31 GC system \citep[see][and references therein]{mackey10,ferguson16}, finding a strong correlation between the spatial position of GCs within the system and stellar streams.

Since not all galaxies have readily visible substructure to compare to GC positions, another approach is to identify deviations in the spatial positions of GCs from the expected symmetric distribution of an unperturbed GC system.
\citet{bonfini12} found an azimuthal asymmetry along the northeast-southwest direction in the GC system of the elliptical galaxy NGC~4261. \citet{dabrusco13} recovered this same feature in NGC~4261 using the K-Nearest Neighbors (KNN) algorithm to estimate the surface density at various points in the field. In doing so, \citet{dabrusco13} probed for radial and azimuthal asymmetries within the NGC~4261 GC system and found that the over-density forms a broken shell or spiral-like pattern. \citet{dabrusco15} probed the cores of the GC systems of the 10 brightest Virgo galaxies for anomalies and found substructures of varying complexity and sizes ($\sim$0.5$\arcsec$ to several arcminutes in length) in all of the observed systems. When studying the Fornax galaxy cluster, \citet{dabrusco16} found an over-density within the NGC~1399 GC system. This over-density is dominated by the blue GCs and stretches from east to west, connecting NGC~1399's GC system to the GC systems of neighboring galaxies. Furthermore, the over-density of blue GCs in Fornax has been shown to be associated with a region of intracluster light within the core of the cluster \citep{iodice17}. \citet{durrell14} and \citet{powalka18} identified GC population features within M87 (NGC~4486) and in the larger Virgo Cluster environment using Kernel Density Estimation (KDE) surface density maps. \citet{durrell14} found the spatial extent of the metal-poor GC populations extends much farther ($\sim$400 kpc from the galaxies) than metal-rich populations around M87 and M84 (NGC~4374), implying that intracluster GC populations are primarily comprised of metal-poor GCs. Dividing their sample into three age bins, \citet{powalka18} detected a spatial over-density of young GCs to the south of M87 that is not present in the intermediate age and oldest GC populations. This over-density was detected at the 5$\sigma$ level and no host galaxy was detected within the over-density region. \citet{madrid18} created a surface density map of the Coma Cluster that highlighted evidence of several ongoing galaxy interactions. Such studies make it clear that GCs can be useful for discovering substructure in galaxy halos and exploring the history and evolution of galaxies and their stellar populations.

In this paper, we make use of existing wide-field CCD imaging data to search for evidence of substructure and galaxy interactions in the GC system of the Virgo Cluster elliptical galaxy M86 (NGC~4406) and the neighboring elliptical galaxy M84  (NGC~4374).  Previous efforts have been made to study the substructure present within the central few arc minutes of the M86 GC system \citep{dabrusco15} and across the Virgo Cluster as a whole \citep{durrell14}. The spatial coverage of our data falls in between these two past works; our images include the two massive elliptical galaxies along with several other low-mass galaxies located at similar distances.  Our study is intended to explore whether the globular cluster systems of the galaxies in the M86 field show any evidence for interactions between the galaxies and their cluster populations.  Theoretical studies predict that environmental effects due to a galaxy cluster tidal field and galaxy interactions can significantly affect the morphological and kinematic properties of GC systems \citep[e.g.][]{ramos15}, strip some GCs from their original host galaxies \citep[e.g][]{forte82,muzzio84,ramos18,ramos20}, and produce intracluster GC populations \citep[e.g.][]{yahagi05,bekki06,bekki09}. The goal here is to carefully search for signatures of these processes in our observational data.

The organization of this paper is as follows.  Section~\ref{sec:obs data} describes our data set and the steps we carried out to produce a list of positions and photometric measurements of GC candidates in the M86/M84 field.  Section~\ref{sec:results} describes both our analysis methods and our results. In this section we report possible substructures within the M86/M84 field, variations in these substructures among the red and blue GC populations, and our findings when studying the kinematics of GCs within the field. The last section of the paper presents a summary of this work.

\section{Observational Data and Globular Cluster Candidate Lists}
\label{sec:obs data}

To investigate substructure in the globular cluster systems of M86 and M84 and the surrounding region, we used the same imaging data that were originally presented in \citet{rz04}.  The goal of the \citet{rz04} study was to quantify the global properties - i.e., the total number of globular clusters, along with the specific frequency, radial distribution, color distribution, and color gradient - of the globular cluster system of M86 and a few other giant early-type galaxies. The images of the M86 field were obtained in March 1999 with the Mosaic imaging camera on the Mayall 4-meter telescope at Kitt Peak National Observatory\footnote{Based on observations at Kitt Peak National Observatory at NSF's NOIRlab, which is managed by the Association of Universities for Research in Astronomy (AURA) under cooperative agreement with the National Science Foundation. The authors are honored to be permitted to conduct astronomical research on Iolkam Du'ag (Kitt Peak), a mountain with particular significance to the Tohono O'odham.}, as part of a wide-field multi-color optical survey of the globular cluster systems of a sample of several spiral, S0, and elliptical galaxies \citep[e.g.,][]{rz01,rz03,rz04, rhode07}. A $\sim$36$\arcmin$ x 36$\arcmin$ area around M86 was
imaged for a total integration time of 3900 sec in $B$, 2700 sec in $V$ and 2100 sec in $R$. A detailed description of how the Mosaic images were processed and calibrated is given in the original \citet{rz04} paper. The pixel scale of the final science images we used is 0.26$\arcsec$~pixel$^{-1}$ and the full width half maximum of the point spread function (FWHMPSF) is
$\sim$0.98$\arcsec$, 1.1$\arcsec$, and 1.2$\arcsec$ in the $B$, $V$, and $R$ images, respectively.

In the original study, \citet{rz04} found that M86 hosts a population of $\sim$2900 globular clusters that extends $\sim$80~kpc from the galaxy center.  The globular cluster system shows at least two peaks in the $B-R$ color distribution as well as a detectable radial color gradient that arises because the red population of globular clusters is more centrally concentrated than the blue population. Later,  \citet{hargis14} used the same Mosaic imaging data to examine the spatial distribution of the globular cluster system of M86 and compare it to the galaxy's light distribution.  They found that the diffuse galaxy light and the globular clusters had similarly flattened, elongated distributions.  The total globular cluster population, as well as the blue and red subpopulations, have nearly identical azimuthal distributions as the galaxy itself, with ellipticities $\epsilon$ $\sim$ 0.4 and position angles of $\sim$$-$60 degrees.

\begin{figure}
    \centering
    \includegraphics[width=.75\columnwidth]{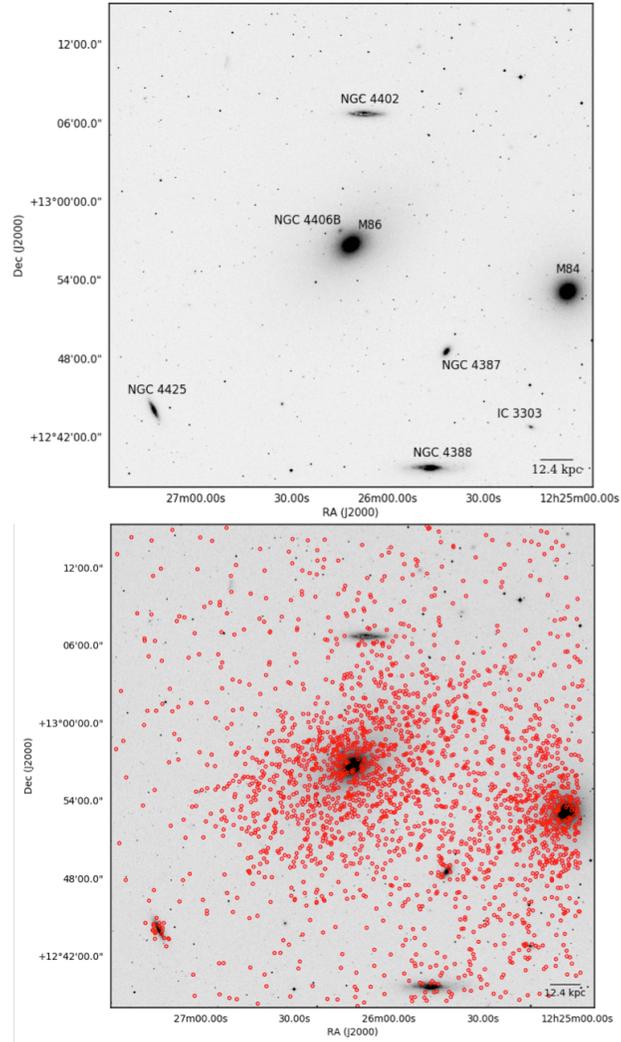}
    \caption{Upper Panel: A Digitized Sky Survey image showing the field-of-view of our Mosaic imaging data, with the eight galaxies in the field labeled.  Our images span a 36$\arcmin$ x 36$\arcmin$ FOV and are centered on the giant elliptical galaxy M86 (NGC 4406). The properties of each of the galaxies in the field are given in Table \ref{table: galaxy properties}. The 2.5$\arcmin$ scale bar corresponds to a length of 12.4 kpc (assuming a distance to the Virgo Cluster of 17 Mpc). Lower Panel: The same field with the spatial positions of the 2250 selected globular cluster candidates (marked as red circles, and selected as described in Section~\ref{sec:obs data} and illustrated in Figure~\ref{fig:bvr_plane}) in our sample.}
    \label{fig:m86_phot}
\end{figure}

Here, we are once again making use of the Mosaic images of the M86 field from \citep{rz04}, but this time our goal is to search for evidence of substructure across the entire field.  Because M86 was the primary target of the original study, it appears at the
center of the Mosaic images, and M84 is positioned on the western
edge (see Figure \ref{fig:m86_phot}).
At the distance of M86 ($\sim$17 Mpc), the 36$\arcmin$ x 36$\arcmin$
field-of-view of the Mosaic images corresponds to a physical area of roughly
180~kpc x 180~kpc. Besides M86 and M84, several other less luminous
Virgo Cluster galaxies appear in the frames.  The basic properties of
the galaxies that appear in the Mosaic images are shown in
Table~\ref{table: galaxy properties} and drawn from the NASA/IPAC Extragalactic Database (NED), unless otherwise specified here or in the table note. The table lists: the galaxy name
(i.e., the NGC number, along with the Messier catalog number and/or
the VCC number); the morphological type from RC3 \citep{devauc91}; the
total absolute magnitude in the $V$-band (calculated by combining
$V{^0_T}$ from RC3 with the distance modulus $m-M$ in table column
(5)); an estimate of the stellar mass of the galaxy, computed by combining the $V$-band absolute magnitude with the appropriate mass-to-light ratio for spiral, S0, or elliptical galaxies, drawn
from \citet{za93}; the galaxy distance modulus and distance in Mpc; and the heliocentric radial velocity of the galaxy. The table also includes (when available) information about the total number of GCs in the system ($N_{\rm GC}$), the specific frequency $S_N$ of the GC system \citep[the number of GCs normalized by the $V$-band absolute magnitude of the galaxy, as defined by][]{harris81}, and $T$, the total number of clusters normalized by the stellar mass of the galaxy \citep[as defined by][]{za93}. As the information in Table~\ref{table: galaxy properties} shows, nearly all of the galaxies that appear in our images have distances within the range $\sim$16 to 18 Mpc; one galaxy, NGC~4387, has a somewhat larger distance (21 Mpc) and another, IC~3303, has no redshift-independent distance measurement in NED.  For the latter, we have assumed a distance modulus of 31.15 (17~Mpc) because the galaxy is included as a likely Virgo Cluster member in \citet{binggeli85}.

\begin{deluxetable*}{lcccccccccccccc}
\centering
\tablecaption{\protect\label{table: galaxy properties}
Properties of the Galaxies in the M86/M84 Images}
\tablehead{
  \\
  Name & Type & $M{_{\rm V}^{\rm T}}$ & log$_{10}$(Mass$_{\rm *}$) &
  $m-M$ & Distance & $V_{r}$ & $N_{\rm GC}$ & $S_N$ & $T$\\
 & &  (mag) & (log($M_{\odot}$)) & (mag) & (Mpc) & (km s$^{-1}$) & & & 
\\
}
\startdata
NGC~4374 (M84, VCC 0763) & E1 & $-$22.3 & 11.8 & 31.32$\pm$0.11 & 18.4 &  1017$\pm$5 & 2800$\pm$200 & 3.5$\pm$0.5 & 4.1$\pm$0.6  \\
NGC~4387 (VCC 0828) & E5 & $-$19.6 & 10.8 & 31.65$\pm$0.73 & 21.4 &  565$\pm$5 & 41$\pm$9 & 0.6$\pm$0.3 & 0.6$\pm$0.3  \\
NGC~4388 (VCC 0836) & Sb & $-$21.7 & 11.4 & 31.29$\pm$0.20 & 18.1 &  2524$\pm$1 & 70$\pm$10 & 0.3$\pm$0.1 & 0.6$\pm$0.3 \\
NGC~4402 (VCC 0873) & Sb & $-$20.0 & 10.7 & 31.02$\pm$0.20 & 16.0 & 232$\pm$5 &  110$\pm$10 &  1.5$\pm$0.3 &  2.2$\pm$0.7  \\
NGC~4406 (M86, VCC 0881) & E3 & $-$22.3 & 11.9 & 31.17$\pm$0.14 & 17.1 & $-$224$\pm$5 & 2900$\pm$400 & 3.5$\pm$0.5 & 4.1$\pm$0.6  \\
NGC~4425 (VCC 0984) & SB0 & $-$19.4 & 10.6 & 31.13$\pm$0.80 & 16.8 & 1908$\pm$5 & 90$\pm$10 & 1.6$\pm$0.5 & 2.5$\pm$0.8 \\
NGC~4406B (VCC 0882) & dE3,N & $-$15.2 & ... & 31.17$\pm$0.14 & 17.1 &  1101$\pm$55 & ... & ... & ... \\
IC~3303 (VCC 0781) & dS0,N & $-$17.5 & ... & ... & ... &  $-$332$\pm$6 & ... & ... & ... \\
\enddata
\tablecomments{For NGC~4402, no $V$-band magnitude is given in RC3 or
  in the NED listing.  Therefore to calculate absolute $V$-band
  magnitude for this galaxy, we have combined $B{^0_T}$ from RC3 with
  an assumed $B-V$ color of 0.65, which is typical for Sb spiral galaxies
  (\citet{rh94}), and the distance modulus in column (5).  
    Distance modulus values for NGC~4374, NGC~4837, and NGC~4406 are from \citet{tonry01}.  Distance modulus values for NGC~4388 and NGC~4402 are from \citet{tully13} and the value for NGC~4425 is from \citet{tully88}. 
  The morphological type for NGC~4406B is taken from \citet{binggeli85}. There is no redshift-independent distance listed in NED for NGC4406B, so we assume the same distance as that of NGC~4406, because these two galaxies are thought to be interacting \citep{elmegreen00}. IC~3303 also has no redshift-independent distance measurement in NED, but because it is identified as a Virgo Cluster member, we have assumed a distance of 17~Mpc ($m-M$ = 31.15).  For both NGC~4406B and IC~3303, we have taken $B{^0_T}$ from \citet{binggeli85} and combined it with the median $B-V$ value of 0.77 for early-type dwarf galaxies from \citet{vanzee04} and the distance modulus to calculate the $M_V^T$ value listed in column (3). The sources for the heliocentric radial velocities listed in column (6) are as follows: the values for  M84, NGC 4387, M86, and NGC 4425 are from \citet{cappellari11}; the value for NGC~4388 is from \citet{lu93}; the value for NGC~4402 is from \citet{binggeli85}; the velocity for NGC~4406B is from \citet{strauss92}; and the velocity for IC~3303 is from \citet{albaretti17}. The GC system properties are drawn from \cite{young16} in all cases except NGC~4406, which are taken from \citet{rz04}, and NGC~4402, which are drawn from Rhode et al. (2020, in preparation).}
\end{deluxetable*}

\begin{figure}
    \centering
    \includegraphics[width=10cm]{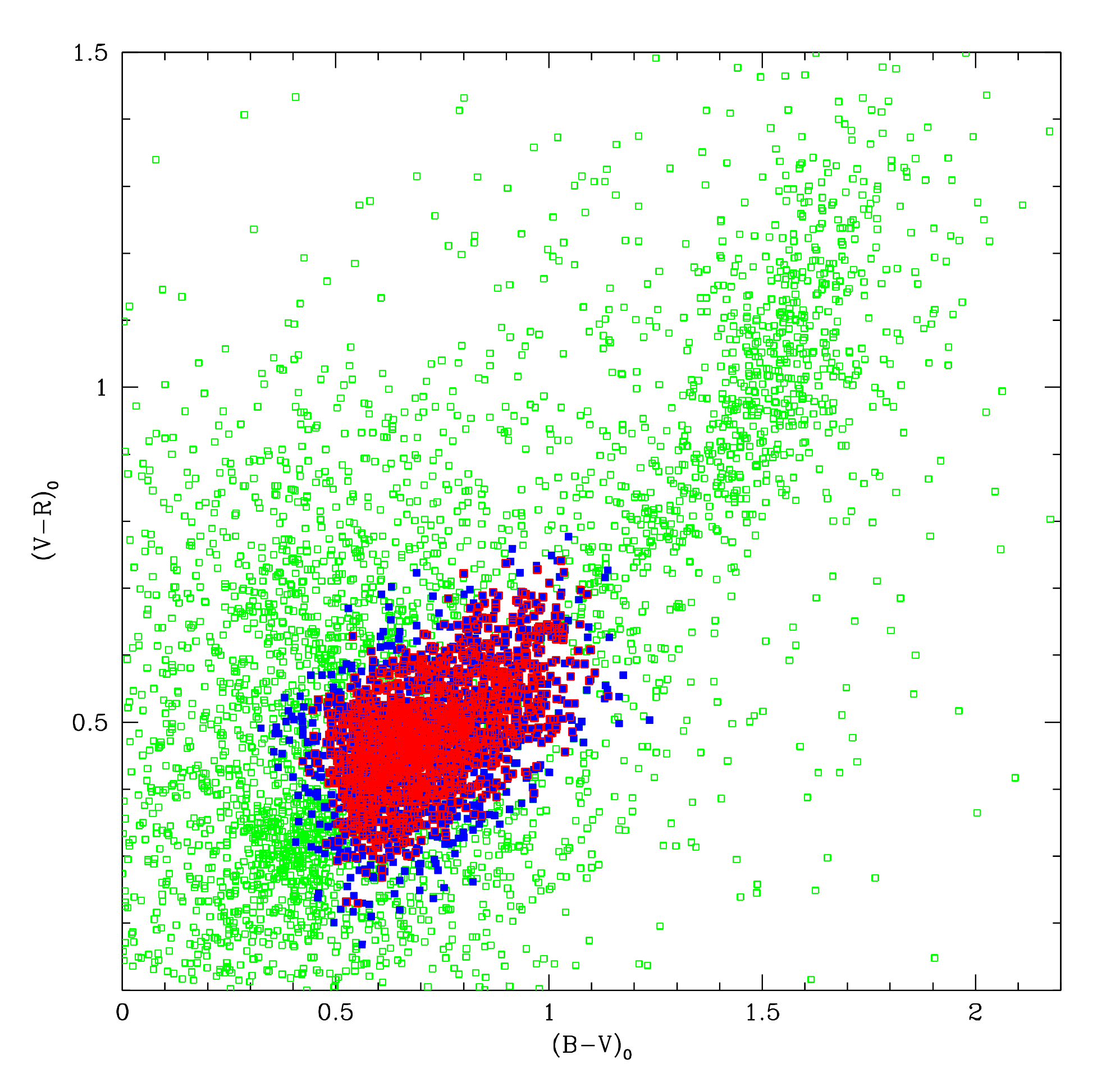}
    \caption{Color selection of globular cluster candidates in the M84/M86 field. The positions in the $V-R$ vs. $B-V$ color-color plane of 5921 point sources that appear in all three filters ($B$, $V$, and $R$) and survived the extended source cut step are plotted as green open squares.  The 2250 globular cluster candidates, which are chosen because they have $V$ magnitudes and $V-R$ vs. $B-V$ colors consistent with their being globular clusters at the distances of the target galaxies, are plotted as blue filled squares. The subset of 1718 globular cluster candidates that satisfy the criteria for the 90\%-complete color sample are plotted on top of the latter, and are shown as red open squares. Details of the globular cluster candidate selection process are given in the text in Section~\ref{sec:obs data}.} 
    \label{fig:bvr_plane}
\end{figure}

For the analysis of M86 that was published in \citet{rz04}, the other
galaxies and globular cluster populations in the Mosaic field were masked out
so that the global properties of M86's globular cluster system (i.e.,
the total number of globular clusters, along with the specific frequency,
radial distribution, color distribution, and color gradient of the
system) could be quantified. For the current analysis, our aim was to 
look for substructure and environmental effects in the globular cluster 
populations in all the galaxies in the field, so we utilized the 
full Mosaic images without masking any large areas. To generate a list of 
globular cluster candidates across
the full field, we carried out the same basic steps that had been used
for the \citet{rz04} analysis.  We began by creating smoothed versions of
the Mosaic images  using a ring filter with a diameter equal to 6 times the FWHMPSF of the images.  We then subtracted the smoothed images from the original images, thereby removing the diffuse starlight associated with the galaxies.  A constant background level was added to the galaxy-subtracted images in order to restore the sky background before the source detection and photometry steps. Next we masked a small central portion of each galaxy (e.g., the central $\sim$200 x 200 pixels in M86 and M84) where the CCD image was saturated and/or where high noise levels in the galaxy-subtracted images prevented the detection of point-source globular cluster candidates. We then used 
IRAF DAOFIND to detect the sources in each image and we matched the source  lists to produce a list of objects that appear in all three filters.   We followed the same procedures that were described and illustrated in \citet{rz04} to remove objects that appear extended in the Mosaic images:  we measured the FWHM of the radial profile of each source and discarded  objects that had larger-than-typical FWHM values for their instrumental magnitudes. At the distances of the Virgo Cluster galaxies in our field, globular clusters will be unresolved in our ground-based images, so objects with large measured FWHM are likely to be faint background galaxies rather than genuine globular clusters.  

Again following the same procedures as were used in the original analysis in \citet{rz04}, we performed aperture photometry on the sources that survived the extended source cut.  We measured the light from each source within a small aperture (with a radius equal to the FWHMPSF value a given image) and then applied the appropriate aperture correction and calibration coefficients to calculate calibrated $V$ magnitudes and $B-V$ and $V-R$ for each source.  The aperture corrections were calculated by measuring the magnitudes of a set of $\sim$20$-$40 bright stars in each image and computing the mean difference between the total magnitude and the magnitude within the small  aperture.  The aperture corrections (total magnitude minus magnitude within the small aperture) ranged from $-$0.15 to $-$0.21 mag with standard deviations of 0.01$-$0.02 mag. The Mosaic images of the M86$+$M84 field were acquired on a night with variable sky conditions, so we calibrated them using $BVR$ exposures of the same field acquired under photometric conditions with the WIYN 3.5-m telescope\footnote{The WIYN Observatory is operated as a joint facility of the NSF’s National
Optical-Infrared Astronomy Research Laboratory, Indiana University,
the University of Wisconsin-Madison, Pennsylvania State University,
the University of Missouri, the University of California-Irvine and
Purdue University.}
and Minimosaic camera. We measured magnitudes and colors for a set of bright stars that appeared in both sets of images and used these values to calculate a set of calibration coefficients (color transformation coefficients and zero-point constants) that could be applied to the Mosaic data. 

We then used our final, photometrically-calibrated point-source photometry to select globular cluster candidates, i.e., unresolved objects with $V$ magnitudes, $B-V$ colors, and $V-R$ colors that are consistent, within the photometric errors of each object, with the magnitudes and colors expected for globular clusters at the distances of the target galaxies. The $BVR$ magnitude and color selection criteria were originally
developed for the globular cluster system survey mentioned earlier in
this section \citep{rz01,rz04}.  For the current analysis of the galaxies that appear in the
M86/M84 field, we selected all point sources in the field with $V$
magnitude greater than 20.0 and colors within a specific region of the  $V-R$ vs. $B-V$ color-color relation defined by Milky Way globular clusters \citep{rz01}. The
$V$ magnitude criterion is based on the assumption that the brightest
globular cluster in the target galaxies will have $M_V$ $\sim$ $-$11
mag.  We note that the galaxies in the M86/M84 field have a range
of distance moduli (from 31.02 to 31.65), so $M_V$ $=$ $-$11 corresponds to a slightly
different apparent $V$ magnitude for each galaxy.  However, because
the $B-V$ and $V-R$ color also play a role in the selection, the final list of globular cluster candidates remains the same
whether we apply the $V$ $>$ 20.0 magnitude criterion across the entire
field, or apply a slightly different $V$ threshold for objects around
each galaxy that takes into account the varying distance moduli of the galaxies.  The $B-V$ and $V-R$ color criteria are explained in detail in \citet{rz01}; briefly, we select objects that have $B-V$ values between 0.56 and 0.99 (which is the $B-V$ range expected for globular clusters with [Fe/H] between 0.0 and $-$2.5, based on the observed colors of Galactic globular clusters) and $V-R$ values that put them within 3 times the standard deviation of the $V-R$ vs. $B-V$ relation for Milky Way globular clusters.  The selection process yielded a final list of 2250 globular cluster
candidates across the entire M86/M84 field. Figure \ref{fig:bvr_plane} shows the globular cluster candidates (blue filled symbols) in the $V-R$ vs. $B-V$ color-color plane, along with the other unresolved sources in the field that were not selected (green open symbols). Note that because the photometric errors for each object are taken into account in the GC candidate selection, the selected candidates (the blue symbols in Figure~\ref{fig:bvr_plane}) extend over a slightly larger region of the color-color plane than they would if the errors were not considered. The spatial locations of the GC candidates in the M86/M84 images are shown in  the bottom panel of Figure \ref{fig:m86_phot}.

Removing extended objects from the source lists and then selecting globular clusters based on their magnitudes and colors in three filters significantly reduces the contamination in our sample \citep[e.g.][]{rz01,rz04}, although some contamination inevitably remains.  Our analysis has shown that many of the contaminating objects that coincide with the globular cluster selection box in the $V-R$ vs. $B-V$ plane are compact background galaxies that are faint enough that they cannot be distinguished from point sources, and some are foreground Galactic stars \citep{rz01,rz04,hargis12}.

To assess the amount of contamination in the globular cluster candidate sample and then correct for it in subsequent steps, we created radial surface density profiles for the globular cluster systems of M84 and M86.  We assigned the globular cluster candidates around each galaxy to a set of 1$\arcmin$-wide annuli, based on the candidates’ projected radial distances from the galaxy center.  We calculated the area of  the portion of each annulus where globular clusters could be detected, i.e., the area minus those parts of the annulus that were masked out (because, for example, they included bright foreground stars or areas with cosmetic defects on the CCD) or extended off the edges of the images.  We then computed the surface density (number per unit area on the sky) of globular cluster candidates for each annulus to construct the radial surface density profile of the system.  The surface density values were highest near the center of each galaxy and then tapered off to a constant, flat value in the outer regions.  We calculated the weighted mean of the surface density in the outer few annuli, where the surface density was constant, and used this as an estimate of the contamination level in the globular cluster candidate sample.  We carried out this analysis for M84 and M86 independently (masking a large region around the neighboring galaxy, and then carrying out the steps to construct the radial profile of the unmasked galaxy), and found that these estimated contamination levels matched each other within the uncertainties.   We also checked to make sure that the contamination correction was the same on the east and west sides of M86, and on the north and south sides of M84.  The estimated contamination level in the globular cluster candidate list for the M84 / M86 field is 0.2997+/-0.0268 per sq. arcmin.  Given the area of the images (1378 square arcminutes), this works out to approximately 413 objects over the entire field, out of the 2,250 objects in the GC candidate list.  

Because we select globular cluster candidates via their $V$ magnitudes and $B-V$ and $V-R$ colors, any contaminants that are present in the globular cluster candidate list will have similar magnitudes and colors as the true globular clusters in the sample. Nevertheless we decided to apply a statistical correction to account for contamination in the globular cluster candidate sample before carrying out our search for spatial features and substructure in the M86/M84 field.  Details about the contamination corrections applied for each part of the analysis are given in Section~\ref{sec:results}.

We carried out a series of artificial star tests in order to quantify
the detection limits for point sources in each of the Mosaic images
and determine how much of the Globular Cluster Luminosity Function
(GCLF) we are able to detect in the target galaxies. We carried out these tests on the $B$, $V$, and $R$ images separately. For each test, we added artificial point sources to a given image, executed the same set of steps we had used to detect the globular cluster candidates, and then determined how many of the artificial sources were recovered.  We added 800 artificial objects to each image, in steps of 0.2 magnitude, and repeated the process until we had covered a magnitude range of 5-6 magnitudes in each filter. The results of the artificial star tests show that our detection process is 50\% complete for point sources with $B=$25.0, $V=$24.0, and $R=$23.4. We can combine the results of the completeness testing
in each of the three filters in order to estimate how much of the
intrinsic GCLF we have detected. Given our point-source detection
limits in $B$, $V$, and $R$, and the requirement that cluster
candidates are detected in all three filters, and assuming an
intrinsic GCLF with a peak magnitude (M$_{\rm V}$ $\sim$ $-$7.4 mag)
and dispersion ($\sigma$ $\sim$ 1.4 mag) \citep[][and references therein]{brodie06} we estimate that we have detected approximately
50\% of the intrinsic GCLF (i.e., the brightest 50\% of the clusters
within the field-of-view of our images) of the two massive ellipticals
M84 and M86.

The numbers above show that the detection completeness varies in each
filter -- the $B$ image is substantially deeper than the $R$ image.
In addition to producing the full list of all 2250 globular cluster
candidates in our images, we also wanted to produce a list that we
could use to explore how certain features in the spatial
distribution of clusters vary when we examine subpopulations of
clusters selected by their colors (see Section \ref{sec:color_maps}).
In other words, we need to construct a sample of globular clusters that is equally complete in all three filters, to be sure that no color selection bias is present in the subsample being used to investigate trends with globular cluster color.  To accomplish this, we defined a sample of objects that we refer to as the ``90\% color sample'', because it is at least 90\% complete in all three filters.  The reddest globular cluster candidate in the full list of 2250 clusters has $B-R$ $=$ 1.87, and in the $B$-band our detection is 90\% complete at 24.77, so a 90\%-complete color sample would include all GC candidates with $R$ brighter than 22.9. Applying this criterion to our GC candidate list yields a sample of 1718 objects. The locations of these 1718 objects in the $V-R$ vs. $B-V$ color-color plane are shown in Figure~\ref{fig:bvr_plane}, plotted in red open symbols on top of the 2250 globular cluster candidates in the full sample.

In the globular cluster system color distributions of elliptical
galaxies, the blue, metal-poor peak often occurs around $B-R$ $\sim$
1.1, the gap between the two populations is typically around $B-R$ $=$
1.23, and the second peak appears at $B-R$ $\sim$ 1.4 
\citep[e.g.][]{rz01,rz04}. Therefore to investigate how the
substructures in the M86/M84 field might vary for the different
globular cluster subpopulations, we divide our 90\% color sample into
two subsamples at $B-R$ $=$ 1.23. We use these blue and red subsamples
in the analysis presented in Section \ref{sec:color_maps}.  We also explain in that section how we used the estimated contamination level (discussed earlier) to apply a statistical correction for contamination to these subsamples.

\section{Analysis and Results} \label{sec:results}
We began our analysis by creating surface density maps of the globular cluster candidates in the M84/M86 field. To look for evidence of substructure, we divided the field into a 150 x 150 grid and used Kernel Density Estimation (e.g. \citealt{silverman86,ivezic14} for a detailed description of the technique) to calculate the surface density maps. Iso-density contours were then added to each map to help emphasize surface density features within the field. We used a Gaussian kernel with a fixed bandwidth of 0.86$\arcmin$ (corresponding to $\sim$200 pixels). This bandwidth was chosen because it appeared to strike the best balance between smoothing away noise within the data while still allowing large-scale substructure to emerge.

\subsection{M86 Surface Density Map}\label{sec:all_KDE_map}
\begin{figure}
    \centering
    \includegraphics[width = .9\columnwidth]{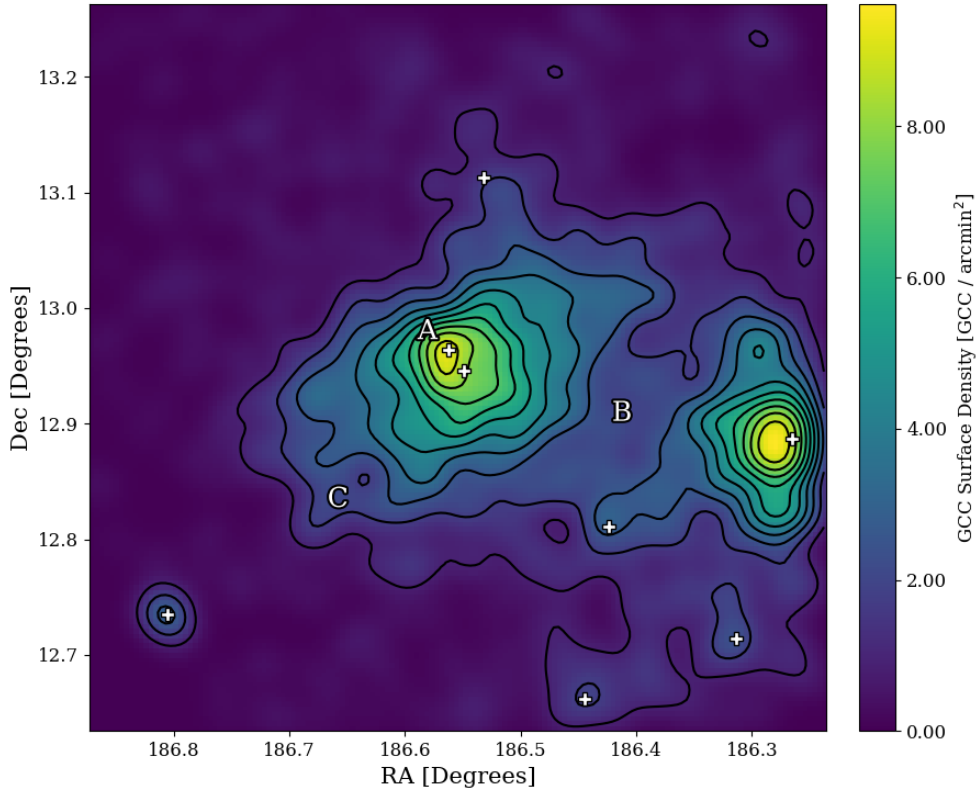}
    \caption{The GC candidate surface density map of the M86 field with contours overlaid. The data were artificially cleaned by randomly selecting and removing the expected number of contaminants from the globular cluster candidate list and performing a KDE analysis on the objects that remained. This was done 100 times with the average density in each grid cell calculated and used to create the final image. The most apparent features are an offset peak in the M86 GC system (A), a bridge connecting M86 to NGC 4387 and M84 (B), and a flattening of the iso-density contours along the southeast side of the M86 GC system (C). The figure is oriented such that north is up and east is to the left and white crosses indicate the positions of galaxies within the field}. 
    \label{fig:all_map}
\end{figure}

Figure \ref{fig:all_map} shows the surface density map for the full GC candidate sample. As explained in the previous section, some of the objects in the sample will not be true globular clusters but will instead be compact objects (foreground stars and background galaxies) that have magnitudes and colors like globular clusters at the distances of the target galaxies, and our sample of 2250 GC candidates should contain approximately 413 of such contaminating objects. To correct for contamination, we artificially cleaned the data by randomly selecting and removing 413 objects from the list of 2250 GC candidates. We then performed the KDE analysis on the cleaned data set.  We carried out these steps a total of 100 times, removing a different random set of 413 objects each time.   From the 100 realizations, we calculated the average surface density for each cell and used those averaged values to create the final KDE map shown in Figure \ref{fig:all_map}. 
We find three major features, marked with an A, B, or C in the figure, along with three minor features.  We discuss each of these features below.

We begin by describing the minor features in the surface density map. First, there appears to be a surface density peak in the M84 GC system located $\sim$1$\arcmin$ to the east of the center of M84. It is important to note that this is an artifact of our method rather than a true surface density anomaly. This offset is a product of edge effects caused by M84's position near the edge of our field. Due to the nature of the KDE technique, surface densities near the edge of the field will be systematically underestimated because of the fact that we cannot account for GCs that are present, but fall outside our field of view.  We varied the kernel size and found the position of this surface density peak largely depended on this parameter, with smaller kernel sizes causing the peak to migrate to the west, toward the center of M84. However, decreasing the kernel size to the degree needed to remove the offset of the surface density peak greatly increased the prominence of noise within the image.

We also see a small connection between the GC systems of M86 and NGC 4402 that appears to indicate a tidal interaction. We used the results from our analysis of NGC~4402’s GC system (see Table~\ref{table: galaxy properties}; Rhode et al.  2020, in preparation) to help us investigate whether this connection might be genuine evidence of a tidal interaction or simply the result of the two GC systems overlapping in projection on the sky.  We removed the globular cluster candidates around NGC~4402 within 2$\arcmin$ (one $R_{\rm eff}$ for the globular cluster system) of the galaxy center; there were 23 candidates inside 2$\arcmin$ in the full sample, and 13 blue candidates and five red candidates in the 90\% color sample. We then repeated the KDE surface density map analysis but with those objects removed. Even with the removal of the GC candidates associated NGC 4402, the connection still appears in all of the KDE surface density maps, although at a reduced strength.

Lastly, in the southeast corner of the field lies NGC 4425, an SB0 galaxy with a remarkably symmetric GC system. While the galaxy is relatively modest in size and mass (with a total magnitude of $M^T_V$ $=$ $-$19.4 and an estimated population of 90$\pm$10 globular clusters; see Table~\ref{table: galaxy properties}), it is interesting that it has survived within the Virgo Cluster without, on first inspection, having had any significant tidal interactions to distort its GC system.

\subsubsection{A Spatial Offset in the Peak Surface Density of the M86 Globular Cluster System}

Feature A in Figure \ref{fig:all_map} is a peak in the GC system surface density that is approximately 0.7$\arcmin$ away from the center of M86. This surface density peak is similar to an over-density reported by \citet{dabrusco15} that lies in roughly the same area. In addition to the unexpected position of the surface density peak, there is a distortion in the central region of the M86 GC system where the surface density contours are elongated along the northeast to southwest direction. 

Since the masked portion of the galaxy (the central $\sim$200~x~200 pixels, which are masked because they are saturated in the CCD images) is close to this surface density peak, we first attempted to determine if this offset might be due to the same kinds of edge effects that created the offset in the M84 surface density peak. Decreasing the kernel size did not cause the peak to migrate towards the center of M86, so we created a 1$\arcmin$ wide annulus just outside and surrounding the masked region and calculated the surface density within this annulus. We then used this surface density along with the area of the masked region to determine the expected number of GC candidates that are likely to be inside the masked region and thus undetected in our images. We randomly distributed the same number of artificial GC candidates within the masked region and used KDE to create a new surface density map. The lopsided morphology of the innermost region of the M86 GC system seen in the original KDE map is still present in this new map.

\begin{figure}
    \centering
    \includegraphics[width = \columnwidth]{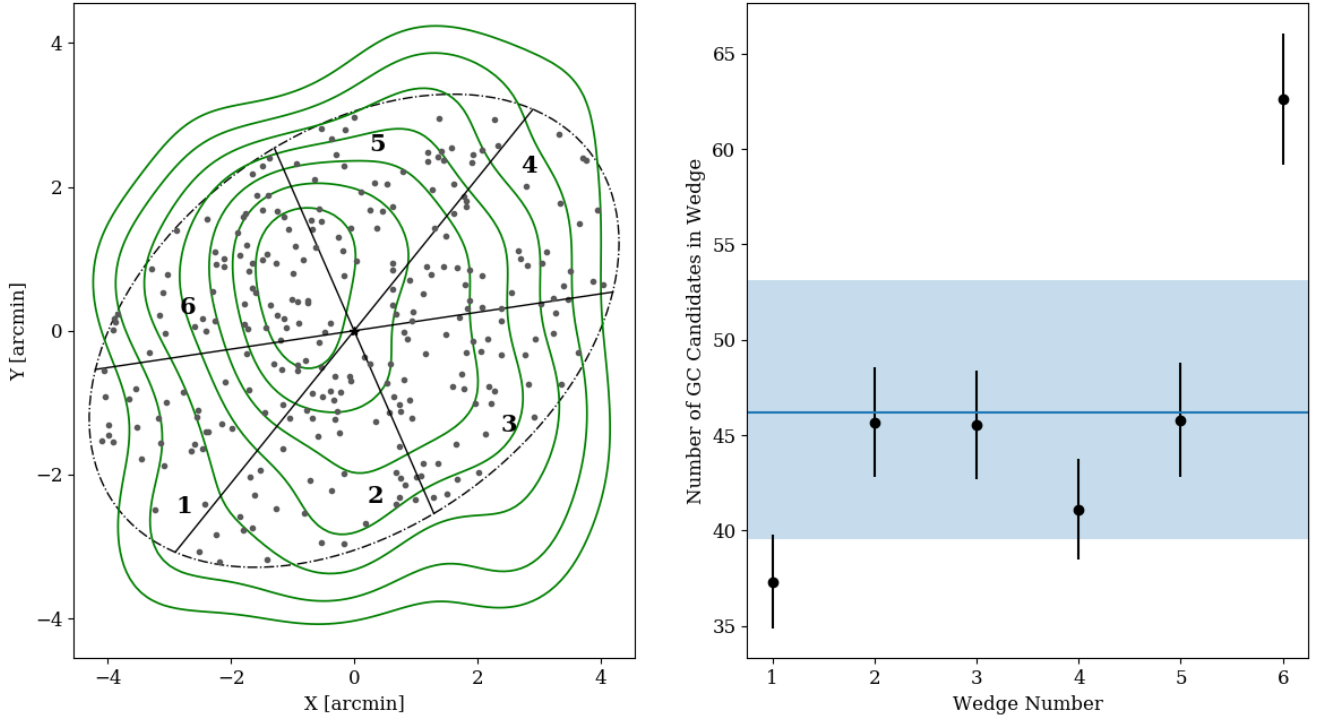}
    \caption{Left panel: The inner 9.2$\arcmin$ x 9.2$\arcmin$ of the M86 GC system. The dotted ellipse outlines the expected morphology of the M86 GC system using characteristics reported in \citet{hargis14} (semi-major axis of 9.2$\arcmin$, ellipticity of 0.38, position angle of -63$^{\circ}$ east of north). The ellipse is split into 6 equal-area, wedge bins and the green lines are contours generated from the KDE measured surface density of the GC candidates within the ellipse. From the contours, it can be seen that the main over-density in the M86 GC system is offset from the center. Wedges are numbered from 1 to 6 for ease of description. Right Panel: The number of GC candidates within each wedge plotted vs. the wedge number. The black points are the number of GC candidates in a given wedge, the blue line is the expected number of GC candidates in each wedge, given no azimuthal asymmetries, and the blue shaded region denotes the area within $\pm$1$\sigma$ Poisson error of the expected number of GC candidates in each wedge.}
    \label{fig:ellipse}
\end{figure}

To quantify the significance of this offset peak, we compared this finding to the expected case of an unperturbed, azimuthally-symmetric GC system. To do this, we first needed to know the parameters that describe the M86 GC system. As mentioned in Section~\ref{sec:obs data}, \citet{hargis14} used the Mosaic images from which our sample was derived to study the spatial distribution of the GC system of M86. They found that the morphology of the inner region of the M86 GC system can be characterized by an ellipse with a semi-major axis length of 9.2$\arcmin$, an ellipticity of 0.38, and a position angle of -63$^{\circ}$ east of north. We created an ellipse with these same characteristics and then divided it into 6 equal-area wedges with the center of the ellipse positioned to coincide with the center of M86. In an unperturbed GC system, we would expect each wedge to contain approximately the same number of GC candidates, but as can be seen in Figure \ref{fig:ellipse}, this is not the case. The left panel in Figure \ref{fig:ellipse} shows our ellipse with the GC candidates and KDE contours plotted; the wedges are numbered and we plot the number of GC candidates in each wedge in the right panel. The horizontal blue line is the expected number of GC candidates in each wedge and the blue-shaded region marks the $\pm$1$\sigma$ Poisson error on that number. The northeast wedge, number 6, has an excess of 20 GC candidates, $\sim$2.5$\sigma$ higher than the expected number, suggesting that this offset GC candidate surface density peak is modestly significant.

It is notable that the dwarf galaxy NGC 4406B (VCC 882) coincides with this over-density. To estimate the number of GCs NGC 4406B might be contributing, we used the data gathered from a survey of dwarf elliptical galaxies by \citet{miller07}. This survey included 13 nucleated dwarf ellipticals with absolute magnitudes between M$_{v}$ = -15 and M$_{v}$ = -16. The estimated number of GCs hosted by these 13 dwarf galaxies ranged from $\sim$2$-$22 clusters. 
Considering this, it is reasonable to assume that NGC 4406B contributes, at least in part, to the over-density seen in Feature A. We compared the color distribution of GC candidates within this over-density to that of the GC candidates in the surrounding area. If NGC 4406B is contributing a significant number of clusters to this over-density, we might expect the color distribution of the GC candidates within the over-density to be shifted towards slightly bluer colors compared to the GC candidates in the surrounding area. For example, \citet{peng06} found that lower-luminosity galaxies have GC systems with bluer mean colors compared to higher-luminosity galaxies, although with plenty of scatter in the relation (see Figure 3 in their paper).

 We found 40 GC candidates within 0.9$\arcmin$ of the over-density and compared their B-R colors to those of the GC candidates in the surrounding regions. We found no evidence to suggest that the GC candidates within this over-density tend to be bluer than the candidates in the surrounding regions, but with such a small sample size and small expected differences in the color distribution this excess is likely to be difficult to detect. It is also possible that this over-density is a result of interactions among the local GCs. \citet{elmegreen00} found that NGC 4406B is undergoing tidal stripping from M86, so a perturbation in the M86 GC system caused by NGC 4406B would potentially explain the existence of this over-density.

\subsubsection{A High Surface Density Bridge Between M86 and M84}
To the southwest of M86, in the area labeled with a B in Figure \ref{fig:all_map}, an overdense region characterized by the presence of a few structural anomalies connects the GC systems of M86 and M84. This region incorporates the GC candidates associated with the moderate-luminosity elliptical galaxy NGC 4387, which has a contour around it that directly connects it to M84. Removal of the GC candidates associated with NGC 4387 does not noticeably affect the appearance of this bridge. The region labeled with a B contains several complex features -- such as the overdense tail emerging to the east of M84 and the general lopsided morphology on the southwest side of M86 -- that are not consistent with the expected spatial distributions of the M86 and M84 GC systems, but instead may be due to an interaction between the two populations.

This region also shares qualitative similarities to structures formed in simulations of galaxy interactions in clusters of galaxies \citep{rudick11,ramos18}, supporting the possibility that it is the result of an interaction between M86 and M84.

\subsubsection{Flattened Contours on the Southeast Side of M86}
The area of the GC system along the southeast edge of M86, labeled with a C in Figure \ref{fig:all_map}, has flattened iso-density contours. This flattening breaks from the expected elliptical shape of the M86 GC system morphology (see Section~\ref{sec:obs data}). We do not see a similar flattening in the contours along the northwest edge of the M86 GC system. This gives an "egg-like" shape to the iso-density contours on the southeast side of M86 and suggests that an accretion event or tidal interaction may have occurred that redistributed the GCs in that region.

\begin{figure}
    \centering
    \includegraphics[width = \columnwidth]{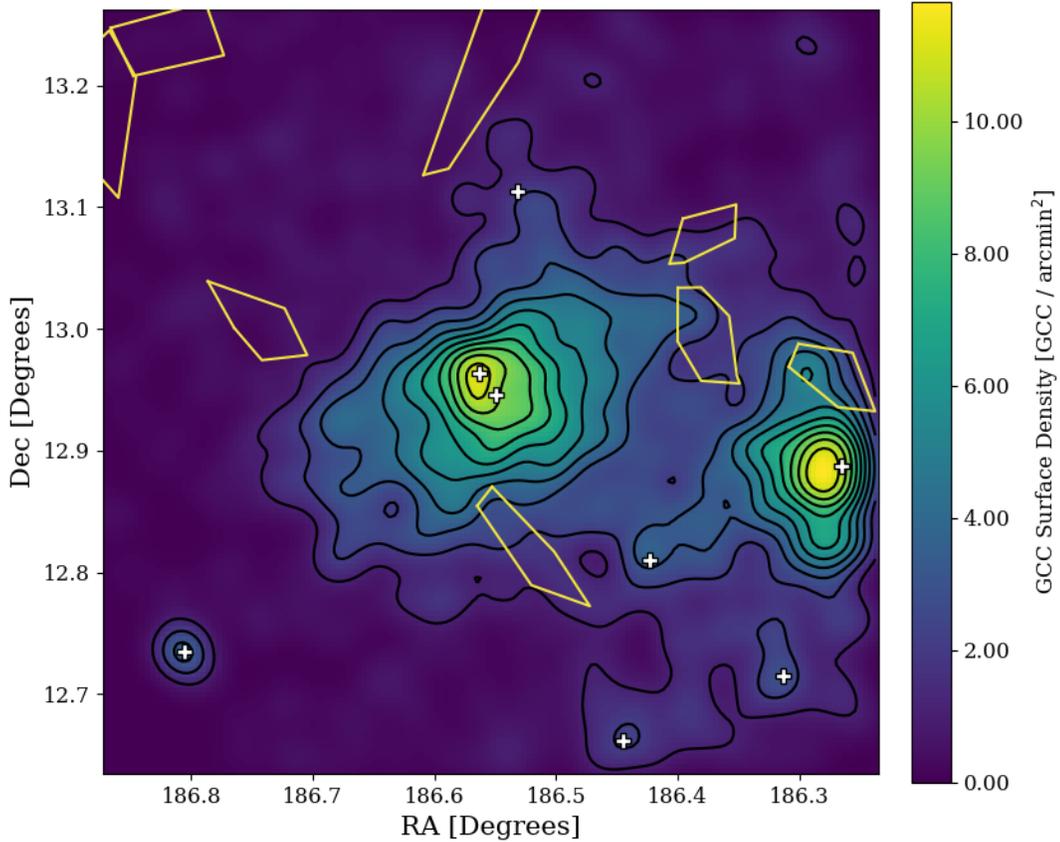}
    \caption{A comparison of our GC candidate surface density map to the regions outlining diffuse light features reported by \citet{janowiecki10}. The yellow boxes correspond to regions surrounding low-surface brightness features. In general, there appears to be little correlation between the substructures seen in the GC candidate surface density and those found using diffuse light in \citet{janowiecki10}. Galaxy positions in the field are marked with  white crosses.}
    \label{fig:regions_map}
\end{figure}

We compared the locations of the features we found in the GC populations to results from two studies that searched for low-surface brightness substructures in this area of the Virgo Cluster. \citet{janowiecki10} used V-band surface photometry to probe the halos of the five brightest ellipticals in the Virgo Cluster for diffuse light features. To observe features nearer the galaxy center, they fitted elliptical isophotes to the galaxy light and subtracted them, leaving behind the diffuse light features. \citet{mihos17} searched the intracluster light between and around galaxies within the Virgo Cluster to detect low-surface brightness features that could indicate past interactions or hierarchical processes. To achieve the depth required to observe features formed from the intracluster light, \citet{mihos17} stacked their images and then removed contamination from Galactic cirrus. Figure \ref{fig:regions_map} is our surface density map with the yellow regions marking features detected in \citet{janowiecki10}. From Figure \ref{fig:regions_map} we can see that the diffuse light structures reported by \citet{janowiecki10} have little to no obvious correlation with the features in our GC candidate surface density map. The region directly to the north of M84 appears to coincide with a small over-density in the surface density map, but because of the proximity to the edge of our field, it is possible that this over-density in the GC system is a spurious result due to edge effects. On the other hand, we do see a similarity between our detected features and a feature that was found in the \citet{mihos17} study of the M86/M84 field. \citet{mihos17} detected an extended low surface brightness feature along the southeast (SE) edge of the M86 halo that did not match the elliptical isophotes and may, they suggested, be a consequence of a major merger event. This substructure is in the same direction as feature C and both are characterized by boxier contours, although the feature we find is comparatively closer in to the galaxy center whereas the "SE Shelf" identified by \citet{mihos17} is formed from diffuse light in the outermost regions of the galaxy.

\subsection{Surface Density Maps of the Red and Blue GC Candidates}\label{sec:color_maps}
\begin{figure}
    \centering
    \includegraphics[width = .8\columnwidth]{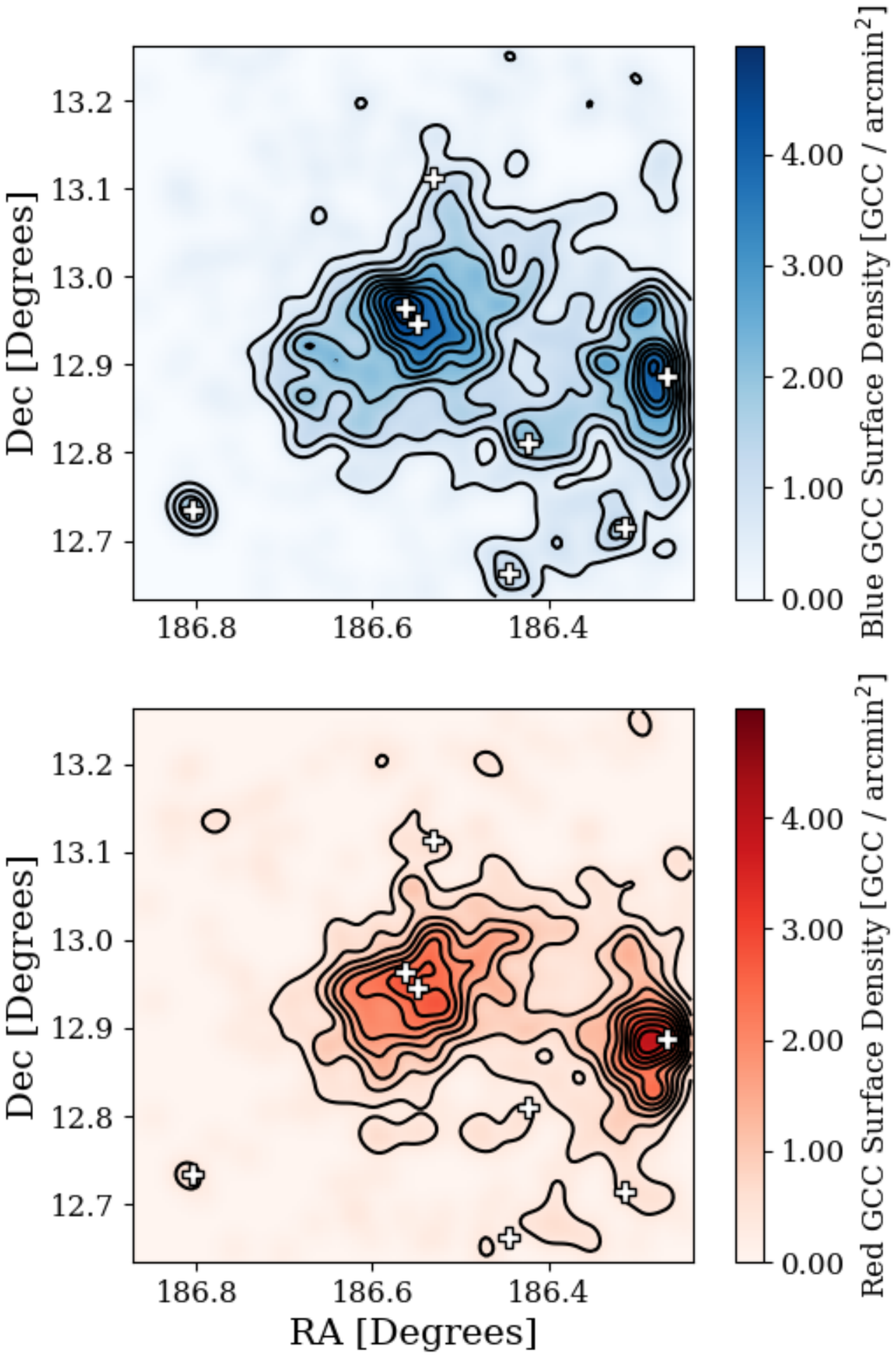}
    \caption{Top Panel: The KDE surface density map of the blue GC candidates in our sample. The high surface density bridge between M86 and M84 has been divided into two bridges, one connecting NGC 4387 to the M84 GC system and one stretching southwest from M86 towards NGC 4387. Bottom Panel: The KDE surface density map of the red GC candidates in our sample. The red GC candidates are more centrally condensed. A small, high-surface density bridge connects the M86 GC system to the M84 GC system, with another connection almost forming to the north of both systems. The asymmetry in the central region of the M86 GC system has disappeared, with the highest surface densities occurring in the middle of the GC system. Both figures are oriented such that north is up and east is to the left. In both panels, galaxy positions are marked with white crosses.}
    \label{fig:color_map}
\end{figure}

Figure~\ref{fig:color_map} shows the surface density maps of the subsamples of GC candidates that are classified as either blue (metal-poor) or red (metal-rich) according to the criteria described in Section \ref{sec:obs data}. To construct this sample and correct it for contamination, we began with the full list of 2250 GC candidates. We randomly removed 413 objects from the sample, and then applied the appropriate $R$ magnitude cut (selecting GC candidates brighter than $R$$=$22.9; see Section~\ref{sec:obs data}) to produce a contamination-corrected 90\% color sample. We then divided the contamination-corrected sample into blue and red subsamples at $B-R$ $=$ 1.23.
As we had done when we constructed the main KDE map shown in Figure \ref{fig:all_map}, we repeated these steps (randomly removing GC candidates to account for contamination, and then applying completeness cuts and splitting the samples according to color) 100 times, removing a different selection of GC candidates each time. In each realization, the surface density was estimated at each cell in our grid and the average surface density for each grid cell was calculated from this collection of estimates. 

The blue GC candidates, shown in the upper panel of Figure~\ref{fig:color_map}, have a more extended spatial distribution compared to their red counterparts. In the surface density map of the blue GC candidates, we see a recurrence of the surface density peak to the northeast of the M86 GC system center that was visible in the map of the full GC candidate sample (feature A from Figure \ref{fig:all_map}). The contours in the inner region of the blue component of the M86 GC system are increasingly elongated along the northeast-southwest axis, suggesting that this elongation comes primarily from the blue GC candidates in the sample. Likewise, the bridge  connecting the M86 and M84 GC systems (feature B from Figure \ref{fig:all_map}) is prominent, with a high surface density arm connecting M84 to NGC 4397 and another high surface density arm emerging from the M86 GC system along the southwest direction. The southeast shelf is still present, but has broken down into patchier, high surface density regions. Furthermore, many of the less luminous galaxies in the field have high-density regions associated with them in the blue GC candidate KDE map, but these high-density regions are much weaker, if they exist at all, in the red GC candidate KDE map. This is not unexpected, because lower-luminosity galaxies tend to have smaller proportions of metal-rich GCs than more luminous galaxies do \citep[e.g.,][]{peng06}.

In contrast to the blue GC candidates, the red GC candidates (lower panel of Figure~\ref{fig:color_map}) are more centrally concentrated. The bridge linking the M86 and M84 GC systems is much weaker, but still present, with a connection formed just north of NGC 4387. The asymmetric surface density peak to the northeast of the M86 GC system center has almost entirely disappeared. The surface density contours in the central region of the red component of the M86 GC system also do not show the same distortion along the northeast-southwest direction that is seen in the blue component, reinforcing that these two features are dominated by metal-poor globular clusters. The southeast shelf is largely non-existent in the surface density map of the red GC candidates. Instead, the central region of the red component of the M86 GC system has boxier iso-density contours than those in the blue GC candidate surface density maps.

\subsection{Spatial Positions of GCs with Velocities}\label{sec:velocity_maps}
Having kinematic data for some of the GC candidates in our sample should provide us with additional insight into the accretion history and ongoing interactions between the GC populations in this well-populated Virgo Cluster field. To further investigate the possibility of substructure in the GC populations of the galaxies, we explored kinematic data published by \citet{park12} and \citet{ko17}. \citet{park12} measured the radial velocities of 25 GCs around M86 and reported a mean GC velocity of $v_{p} = -354^{+81}_{-79}$ km s$^{-1}$ with a velocity dispersion of $\sigma_{p}=292\pm32$ km s$^{-1}$. \citet{ko17} carried out a wide-field spectroscopic survey of GCs within the Virgo Cluster. Of the 201 GC spectra in the \citet{ko17} sample, 94 GCs were within our field, with one of these GCs also appearing in the \citet{park12} sample. After combining the \citet{park12} and \citet{ko17} data sets and removing the one duplicate GC, we had a total of 118 GCs with radial velocities in our field to study.

\begin{figure}
    \centering
    \includegraphics[width=.7\columnwidth]{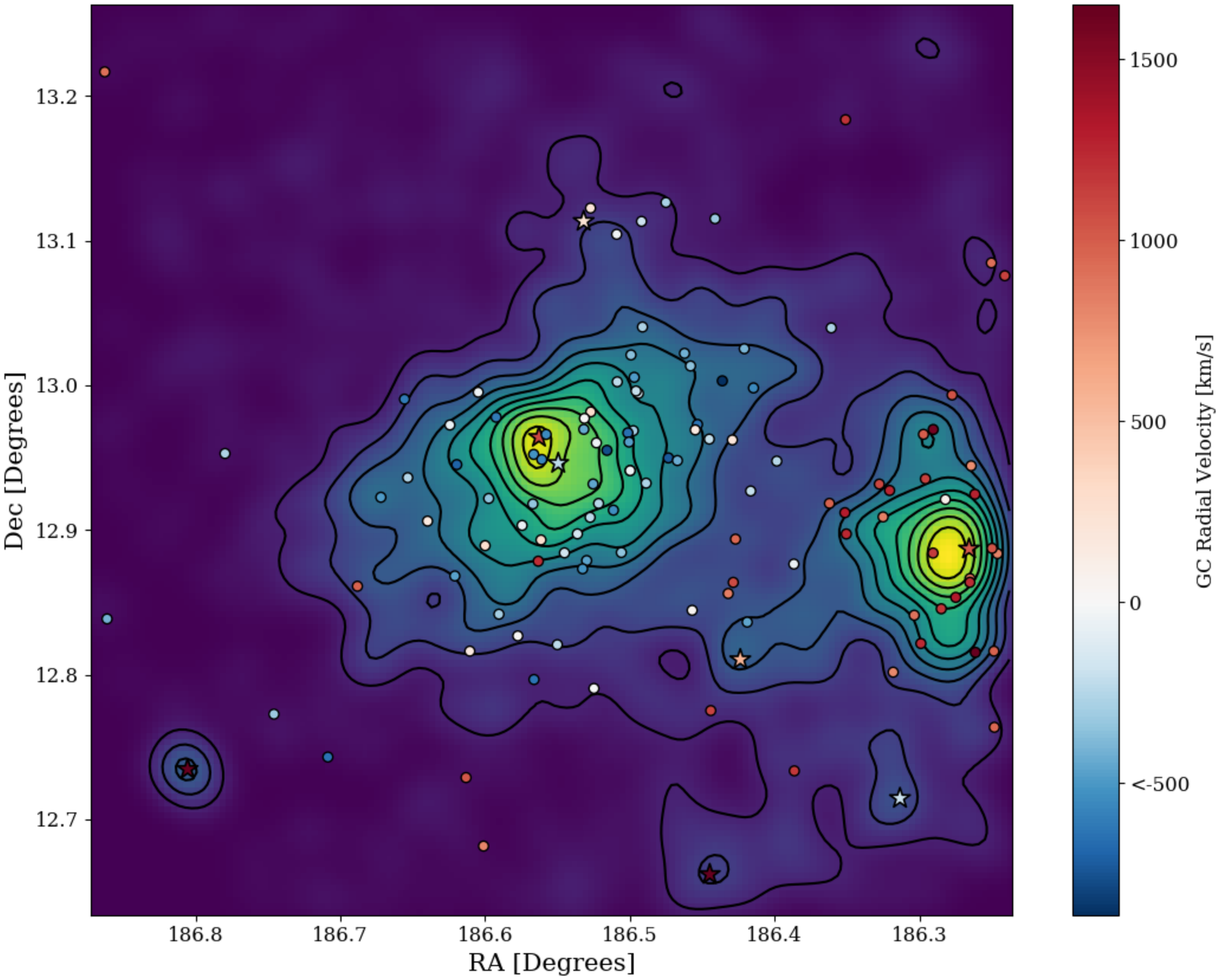}
    \caption{The spatial distribution of GCs with measured radial velocities in the region around M86 (left) and M84 (right). The GCs are plotted on top of the contamination-corrected KDE map of our full GC candidate sample. Each GC with a measured velocity is color-coded according to its reported radial velocity, with red GCs receding and blue GCs approaching. Each galaxy's position is marked with a star with the color indicating that galaxy's radial velocity.}
    \label{fig:vel_map}
\end{figure}

Figure \ref{fig:vel_map} shows a map of the spatial distributions of the collated sample of GCs within our field from the \citet{park12} and \citet{ko17} surveys plotted on top of the surface density map of our full, contamination-corrected GC candidate sample. The points representing the GCs are color-coded according to their radial velocity in km s$^{-1}$ and the spatial positions of galaxies within the field are denoted by stars, also color-coded according to their radial velocities as reported in Table \ref{table: galaxy properties}. From this figure we see that GCs with measured velocities populate all of the major features discussed previously. The GC candidate surface density peak to the northeast of M86's spatial position has three clusters with negative radial velocities, consistent with the velocity of M86 (-224$\pm 5$ km s$^{-1}$). The bridge of elevated GC candidate surface density that connects the M86 and M84 GC systems is populated by several GCs with varying radial velocities, suggesting it is comprised of GCs from both galaxies and, possibly, intracluster GCs. The flattened shelf along the southeast side of the M86 GC system is mostly populated by GCs with velocities consistent with the radial velocity of M86, but there are two GCs with high recession velocities more akin to the radial velocities of M84 (1017$\pm$5 km s$^{-1}$) or NGC 4406B (1101$\pm$55 km s$^{-1}$).

Most of the less massive galaxies in the field appear to host no, or very few, of the globular clusters with measured velocities. The barred lenticular galaxy NGC 4425 and the spiral galaxy NGC 4388 each have two clusters in close proximity with recession velocities similar to those of the galaxies, suggesting that the clusters are associated. NGC 4402 also has two clusters with similar velocities nearby, but its close proximity to M86 makes it difficult to assess to which galaxy's GC system these clusters belong. 

\begin{figure}
    \centering
    \includegraphics[width=.7\columnwidth]{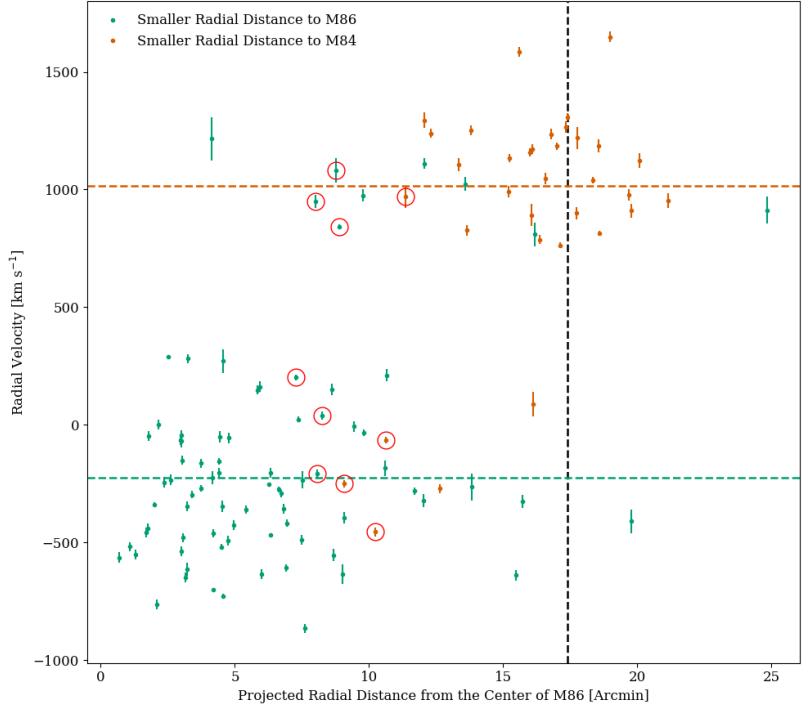}
    \caption{The radial velocities of GCs from the collated survey data of \citet{park12} and \citet{ko17} as a function of each object's projected radial distance from the center of M86. Points have been color-coded based on the closest massive elliptical galaxy, with green (orange) points having a smaller projected radial distance from M86 (M84). The three dashed lines are color-coded as follows: green represents the radial velocity of M86, orange represents the radial velocity of M84, and black represents the projected radial distance from the center of M84 to the center of M86. All data points have uncertainties associated with their measured velocities, but many of these uncertainties are smaller than the markers. The ten GC candidates located within the bridge between M86 and M84 are circled.}
    \label{fig:rad_vel}
\end{figure}

We plot the radial velocity of these globular clusters as a function of their projected radial distance from the center of M86 in Figure \ref{fig:rad_vel}. The clusters have been color-coded based on their proximity to either M86 or M84, with green points representing clusters with a smaller projected radial distance to the center of M86 and orange points representing clusters with a smaller distance to the center of M84. The radial velocities of M86 and M84 (as reported by \citealt{cappellari11}) are represented as the green and orange dashed lines, respectively, with the vertical, black dashed line denoting the projected radial distance between M86 and M84. There is clearly a large gap between $\sim$300 km s$^{-1}$ and $\sim$800 km s$^{-1}$ where GCs are lacking; the large kinematic separation could be indicative of a fly-by interaction between M86 and M84. This plot also confirms the existence of several clusters with kinematics that are not consistent with the expected velocity based on their closest massive elliptical neighbor.

We attempted to quantify the significance of differences between the radial velocities of 114 GCs in this sample with the radial velocities of each GC's closest massive elliptical galaxy. We removed both sets of clusters nearest NGC~4425 and NGC~4388, as these GCs have radial velocities in close agreement with those of NGC~4425 and NGC~4388, and are unlikely to be hosted by either M86 or M84. To determine whether a GC had a radial velocity that was significantly different than the mean velocity of the nearest massive elliptical galaxy, we needed to estimate the velocity dispersion of the M86 and M84 GC systems. We did this using the biweight method described in \citet{beers90} with GCs within one effective radius of M86 and GCs within one effective radius of M84. We limited ourselves to GCs within one effective radius of each galaxy because of the significant overlap between the two systems. With this method we report a velocity dispersion of $\sigma_{v,M86} = 308^{+60}_{-49}$ km s$^{-1}$ and $\sigma_{v,M84} = 265^{+83}_{-62}$ km s$^{-1}$ calculated from 46 and 25 GCs, respectively.

\begin{figure}
    \centering
    \includegraphics[width=.7\columnwidth]{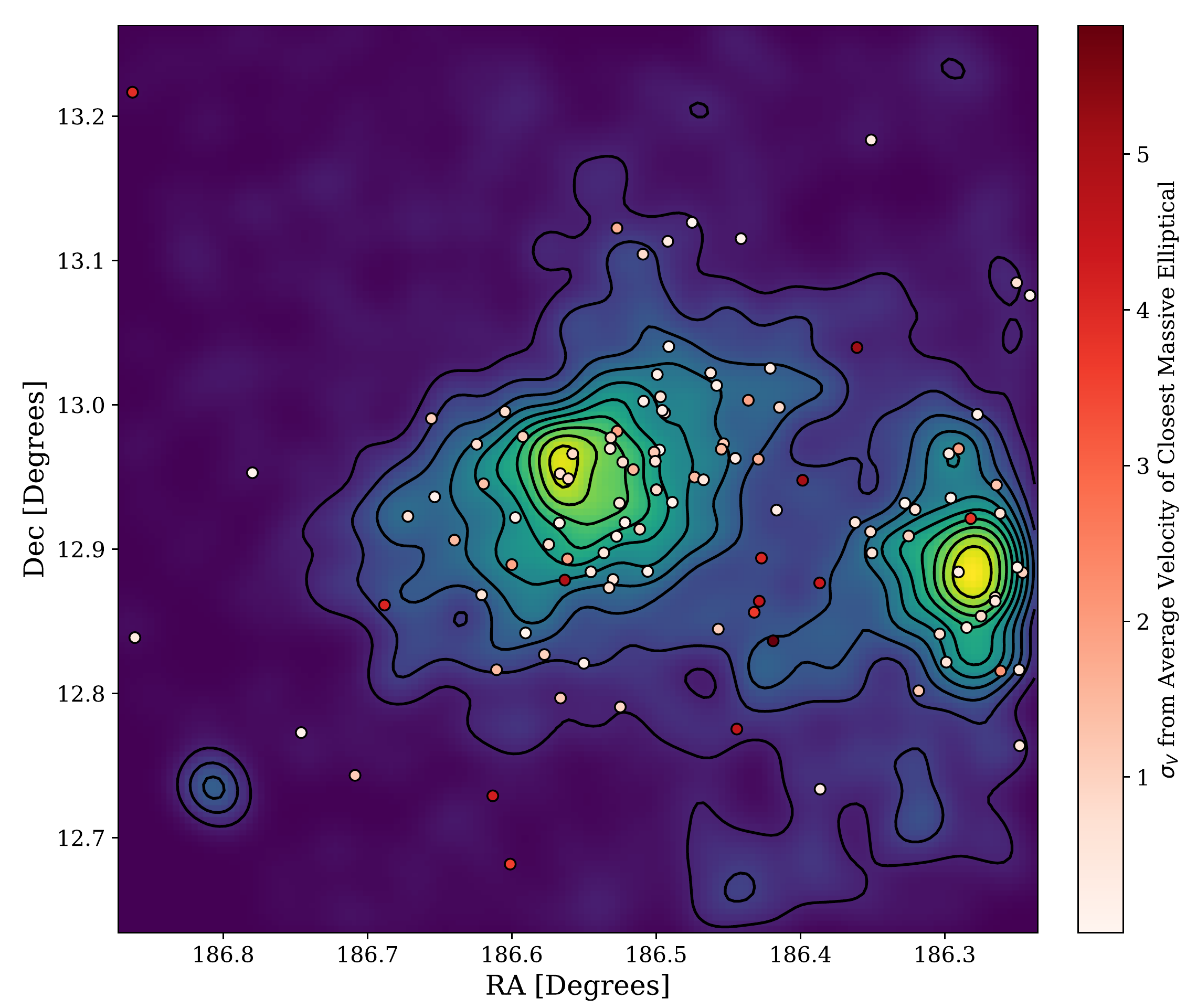}
    \caption{The spatial positions of the GCs with reported velocities, color-coded based on the difference between the GC velocity and the average velocity of the nearest massive elliptical galaxy. Most of the GCs with radial velocities inconsistent with their closest massive elliptical galaxy are located within the bridge and stretch along the east-west axis and to the south of the M86 GC system.}
    \label{fig:dispersions}
\end{figure}

In Figure \ref{fig:dispersions}, we color-code each GC based on the difference between the cluster's velocity and the average velocity of the closest massive elliptical galaxy (normalized to the velocity dispersion of the closest massive elliptical galaxy). The bridge connecting the GC systems of M86 and M84 hosts ten GCs. The GCs within this connection form a rough line from the northwest to the southeast, across the feature itself. These GCs have a range of velocities, from -454 km s$^{-1}$ to +1081 km s$^{-1}$, and all of them have velocities consistent with those expected for GCs associated with either M86 or M84. However, as shown in Figure \ref{fig:dispersions}, six of these GCs have radial velocities that are 3$\sigma$ to 6$\sigma$ different from the radial velocity of their closest massive elliptical neighbor. Several of the clusters in the field surrounding the two galaxies, along with two clusters located between NGC~4425 and NGC~4388, have velocities consistent with the GCs of M84.
Similarly, there are a few clusters that are located closer to the center of M84 but have velocities consistent with that of M86. Although some of the clusters in question may have been stripped by tidal interactions between the two massive ellipticals, they may also simply represent clusters in the outer regions of the globular cluster systems of M84 and M86. A significantly larger sample of velocities would be needed to construct a complete characterization of the phase space of these globular cluster systems, to explore the possible fingerprints of the tidal interaction between M84 and M86, and to identify a population of intracluster globular clusters \citep[see, e.g.,][]{romanowsky12,longobardi15,longobardi18}.

\section{Conclusions}\label{sec:conc}
In this paper we presented results from our search for substructure within the GC populations of M86 and M84 using the spatial positions of 2250 GC candidates selected from wide-field, multi-color broadband imaging. We estimated the amount of contamination in the GC candidate sample and then applied statistical corrections to account for it. We then used the KDE technique to create surface density maps of the GC systems within the M86/M84 field and of the red and blue GC candidate subpopulations separately. We combined our KDE analysis results with kinematic data drawn from \citet{park12} and \citet{ko17} and analyzed the spatial positions and radial velocities of a subset of the GCs to see whether the GC kinematics correlate with the features we detect. Our results can be summarized as follows:
\newline

\hangindent=.7cm 1. We identify three main substructures within the field. First, we find the surface density peak of the M86 GC system is offset from the system's center by approximately 0.7$\arcmin$. This over-density coincides with the dwarf elliptical galaxy, NGC 4406B. Although the GC system of NGC 4406B has not been studied, dwarf elliptical galaxies with similar luminosities can host enough GCs to account, at least partially, for this over-density. Moreover, \citet{elmegreen00} has found evidence for an interaction between M86 and NGC 4406B, so a perturbation within the M86 GC system caused by NGC 4406B is within the realm of possibility. The second structure we detect is a bridge connecting the M86 and M84 GC systems, perhaps as a result of interactions between the two galaxies. The final structure is a flattened iso-density contour along the southeast edge of the M86 GC system. \citet{mihos17} found a similar structure in the low surface brightness starlight around M86 and called it the "SE Shelf", arguing that it may be the result of a major accretion event in the galaxy's past.
\newline

\hangindent=.7cm 2. When we divide our GC candidates into red (metal-rich) and blue (metal-poor) subpopulations we find that the blue GC candidates are the dominant population in the structures mentioned above: the various features that are apparent when we examine the total GC candidate sample (the surface density peak offset from the M86 GC system center, the bridge connecting the M86 and M84 GC systems, the flattened iso-density contour along the southeast of the M86 GC system) are also apparent when we observe the blue GC candidate subsample. In contrast, these features have a marked decrease in strength when analyzing the surface density map of the red GC candidate sample.
\newline

\hangindent=.7cm 3. We find several examples of GCs with radial velocities that are inconsistent with the velocity of the nearest giant elliptical galaxy. The majority of such GCs are located within the bridge connecting the M86 and M84 GC systems. It is possible that the GCs within this bridge have been stripped from their host galaxy, but they may instead simply be part of the outer GC population of these two massive elliptical galaxies. Our analysis also shows possible evidence for a stream of intracluster GCs with velocities consistent with that of M84 emerging along the east-west axis of M86 and to the south of the M86 GC system.

\acknowledgments

The research described in this paper was supported in part by a Daniel Kirkwood Fellowship from the Indiana University Astronomy Department to RAL. We are grateful to the anonymous referee for their careful review of the manuscript and valuable suggestions that improved the paper. This research has made use of the NASA/IPAC Extragalactic Database (NED) which is operated by the Jet Populsion Laboratory, California Institute of Technology, under contract with the National Aeronautics and Space Administration. This research made use of Astropy, \footnote{http://www.astropy.org} a community-developed core Python package for Astronomy \citep{astropy:2013, astropy:2018}.


\begin{thebibliography}{}

\bibitem[Albaretti et al.(2017)]{albaretti17} Albaretti, F.D., et al. 2017, \apjs, 233, 25
\bibitem[Armandroff \& Zinn (1988)]{az88} Armandroff, T.E. \& Zinn, R. 1988, \aj, 96, 92 
\bibitem[Ashman \& Zepf (1992)]{az92} Ashman, K.M. \& Zepf, S.E. 1992,
  \apj, 384, 50
\bibitem[Astropy Collaboration et al.(2013)]{astropy:2013} Astropy Collaboration, Robitaille, T.~P., Tollerud, E.~J., et al.\ 2013, \aap, 558, A33
\bibitem[Astropy Collaboration et al.(2018)]{astropy:2018} Astropy Collaboration, Price-Whelan, A.~M., Sip{\H{o}}cz, B.~M., et al.\ 2018, \aj, 156, 123
\bibitem[Beers et al.(1990)]{beers90} Beers, T.~C., Flynn, K., \& Gebhardt, K.\ 1990, \aj, 100, 32
\bibitem[Bekki \& Yahagi(2006)]{bekki06} Bekki, K., \& Yahagi, H.\ 2006, \mnras, 372, 1019
\bibitem[Bekki \& Yahagi(2009)]{bekki09} Bekki, K., \& Yahagi, H.\ 2009, Globular Clusters - Guides to Galaxies, 373
\bibitem[Belokurov et al.(2006)]{belokurov06} Belokurov, V., Zucker, D.~B., Evans, N.~W., et al.\ 2006, \apjl, 642, L137
\bibitem[Belokurov et al.(2018)]{belokurov18} Belokurov, V., Erkal, D., Evans, N.~W., et al.\ 2018, \mnras, 478, 611
\bibitem[Binggeli et al. (1985)]{binggeli85} Binggeli, B., Sandage, A., \& Tammann, G. 1985, \aj, 90, 1681
\bibitem[Blakeslee(1999)]{blakeslee99} Blakeslee, J.~P.\ 1999, \aj, 118, 1506
\bibitem[Bonfini et al.(2012)]{bonfini12} Bonfini, P., Zezas, A., Birkinshaw, M., et al.\ 2012, \mnras, 421, 2872
\bibitem[Brodie \& Strader (2006)]{brodie06} Brodie, J.P. \& Strader,
  J. 2006, \araa, 44, 193  
\bibitem[Cappellari et al.(2011)]{cappellari11} Cappellari, M., Emsellem, E., Krajnovi{\'c}, D., et al.\ 2011, \mnras, 413, 813
\bibitem[C{\^o}t{\'e}(1999)]{cote98} C{\^o}t{\'e}, P.\ 1999, \aj, 118, 406
\bibitem[Crnojevic et al. (2016)]{crnojevic16} Crnojevic, D., Sand, D.J., Spekkens, K., Caldwell, N., Guhathakurta, P., McLeod, B., Seth, A., Simon, J.D., Strader, J., \& Toloba, E. 2016, \apj, 823, 19
\bibitem[D'Abrusco et al.(2013)]{dabrusco13} D'Abrusco, R., Fabbiano, G., Strader, J., et al.\ 2013, \apj, 773, 87
\bibitem[D'Abrusco et al.\ (2015)]{dabrusco15} D'Abrusco, R., Fabbiano, G., \& Zezas, A., 2015, \apj, 805, 26
\bibitem[D'Abrusco et al.(2016)]{dabrusco16} D'Abrusco, R., Cantiello, M., Paolillo, M., et al.\ 2016, \apjl, 819, L31
\bibitem[de Vaucouleurs et al.\ (1991)]{devauc91} de~Vaucouleurs, G.,
de~Vaucouleurs, A., Corwin Jr., H.G., Buta, R.J., Paturel, G., \&
Fouque, P. 1991, Third Reference Catalogue of Bright Galaxies (New
York: Springer)
\bibitem[Deason et al.(2018)]{deason18} Deason, A.~J., Belokurov, V., Koposov, S.~E., et al.\ 2018, \apjl, 862, L1
\bibitem[Durrell et al.\ (2014)]{durrell14} Durrell, P.~R., C{\^o}t{\'e}, P., Peng, E.~W., et al.\ 2014, \apj, 794, 103
\bibitem[El-Badry et al.(2019)]{elbadry19} El-Badry, K., Quataert, E., Weisz, D.~R., et al.\ 2019, \mnras, 482, 4528
\bibitem[Elmegreen et al.\ (2000)]{elmegreen00} Elmegreen, D.~M., Elmegreen, B.~G., Chromey, F.~R., \& Fine, M.~S.\ 2000, \aj, 120, 733
\bibitem[Ferguson et al.(2002)]{ferguson02} Ferguson, A.~M.~N., Irwin, M.~J., Ibata, R.~A., et al.\ 2002, \aj, 124, 1452
\bibitem[Ferguson \& Mackey(2016)]{ferguson16} Ferguson, A.~M.~N., \& Mackey, A.~D.\ 2016, Tidal Streams in the Local Group and Beyond, 191
\bibitem[Forbes \& Bridges(2010)]{forbes10} Forbes, D.~A., \& Bridges, T.\ 2010, \mnras, 404, 1203
\bibitem[Forte et al.\ (1982)]{forte82} Forte, J.C., Martinez, R.E.,
  \& Muzzio, J.C. 1982, \aj, 87, 1465
\bibitem[Gaia Collaboration et al.(2016)]{gaia16} Gaia Collaboration, Prusti, T., de Bruijne, J.~H.~J., et al.\ 2016, \aap, 595, A1
\bibitem[Hargis \& Rhode(2012)]{hargis12} Hargis, J.~R., \& Rhode, K.~L.\ 2012, \aj, 144, 164
\bibitem[Hargis \& Rhode(2014)]{hargis14} Hargis, J.~R., \& Rhode, K.~L.\ 2014, \apj, 796, 62
\bibitem[Harris \& van den Bergh (1981)]{harris81} Harris, W.E. \& van den Bergh, S. 1981, \aj, 86, 1627
\bibitem[Helmi et al.(2018)]{helmi18} Helmi, A., Babusiaux, C., Koppelman, H.~H., et al.\ 2018, \nat, 563, 85
\bibitem[Hopkins et al.(2014)]{hopkins14} Hopkins, P.~F., Kere{\v{s}}, D., O{\~n}orbe, J., et al.\ 2014, \mnras, 445, 581
\bibitem[Hopkins et al.(2018)]{hopkins18} Hopkins, P.~F., Wetzel, A., Kere{\v{s}}, D., et al.\ 2018, \mnras, 480, 800
\bibitem[Ibata et al.(1994)]{ibata94} Ibata, R.~A., Gilmore, G., \& Irwin, M.~J.\ 1994, \nat, 370, 194
\bibitem[Ibata et al.(2001)]{ibata01} Ibata, R., Irwin, M., Lewis, G., et al.\ 2001, \nat, 412, 49
\bibitem[Iodice et al.(2017)]{iodice17} Iodice, E., Spavone, M., Cantiello, M., et al.\ 2017, \apj, 851, 75
\bibitem[Ivezi\'c et al. (2014)]{ivezic14} Ivezi\'c, \v Z., Connolly, A., VanderPlas, J., Gray, A. \ 2014, Statistics, Data Mining, and Machine Learning in Astronomy, (Princeton University Press)
\bibitem[Janowiecki et al.(2010)]{janowiecki10} Janowiecki, S., Mihos, J.~C., Harding, P., et al.\ 2010, \apj, 715, 972
\bibitem[Ko et al.(2017)]{ko17} Ko, Y., Hwang, H.~S., Lee, M.~G., et al.\ 2017, \apj, 835, 212
\bibitem[Li \& Gnedin(2014)]{li14} Li, H., \& Gnedin, O.~Y.\ 2014, \apj, 796, 10
\bibitem[Lim et al.\ (2017)]{lim17} Lim, S., Peng, E.~W., Duc, P.-A., et al.\ 2017, \apj, 835, 123
\bibitem[Longobardi et al.(2015)]{longobardi15} Longobardi, A., Arnaboldi, M., Gerhard, O., et al.\ 2015, \aap, 579, A135
\bibitem[Longobardi et al.(2018)]{longobardi18} Longobardi, A., Peng, E.~W., C{\^o}t{\'e}, P., et al.\ 2018, \apj, 864, 36
\bibitem[Lu et al.(1993)]{lu93} Lu, N.~Y., Hoffman, G.~L., Groff, T., et al.\ 1993, \apjs, 88, 383
\bibitem[Mackey et al.(2010)]{mackey10} Mackey, A.~D., Huxor, A.~P., Ferguson, A.~M.~N., et al.\ 2010, \apj, 717, L11
\bibitem[Madrid et al.\ (2018)]{madrid18} Madrid, J.~P., O'Neill, C.~R., Gagliano, A.~T., et al.\ 2018, \apj, 867, 144
\bibitem[Mieske et al.(2004)]{mieske04} Mieske, S., Infante, L., Ben{\'\i}tez, N., et al.\ 2004, \aj, 128, 1529
\bibitem[Mihos et al.(2017)]{mihos17} Mihos, J.~C., Harding, P., Feldmeier, J.~J., et al.\ 2017, \apj, 834, 16
\bibitem[Miller \& Lotz(2007)]{miller07} Miller, B.~W., \& Lotz, J.~M.\ 2007, \apj, 670, 1074
\bibitem[Muzzio et al.\ (1984)]{muzzio84} Muzzio, J.C., Martinez,
  R.E., \& Rabolli, M. 1984, \apj, 285, 7  
\bibitem[Myeong et al.(2018)]{myeong18} Myeong, G.~C., Evans, N.~W., Belokurov, V., et al.\ 2018, \apjl, 863, L28
\bibitem[Park et al.(2012)]{park12} Park, H.~S., Lee, M.~G., \& Hwang, H.~S.\ 2012, \apj, 757, 184
\bibitem[Peng et al.(2006)]{peng06} Peng, E.~W., Jord{\'a}n, A., C{\^o}t{\'e}, P., et al.\ 2006, \apj, 639, 95
\bibitem[Powalka et al.(2018)]{powalka18} Powalka, M., Puzia, T.~H., Lan{\c{c}}on, A., et al.\ 2018, \apj, 856, 84
\bibitem[Ramos et al.(2015)]{ramos15} Ramos, F., Coenda, V., Muriel, H., et al.\ 2015, \apj, 806, 242
\bibitem[Ramos-Almendares et al.(2018)]{ramos18} Ramos-Almendares, F., Abadi, M., Muriel, H., et al.\ 2018, \apj, 853, 91
\bibitem[Ramos-Almendares et al.(2020)]{ramos20} Ramos-Almendares, F., Sales, L.~V., Abadi, M.~G., et al.\ 2020, In Press, arXiv:1906.11921 [astro-ph.GA]
\bibitem[Rhode(2012)]{rhode12} Rhode, K.~L.\ 2012, \aj, 144, 154
\bibitem[Rhode \& Zepf (2001)]{rz01} Rhode, K.L. \& Zepf, S.E.\ 2001,
  \aj, 121, 210
\bibitem[Rhode \& Zepf (2003)]{rz03} Rhode, K.L. \& Zepf, S.E. \ 2003, \aj, 126, 2307
\bibitem[Rhode \& Zepf (2004)]{rz04} Rhode, K.L. \& Zepf, S.E. \ 2004,
  \aj, 127, 302
\bibitem[Rhode et al. (2007)]{rhode07} Rhode, K.L., Zepf, S.E., Kundu, A., \& Larner, A.N. \ 2007, 134, 1403
\bibitem[Roberts \& Haynes (1994)]{rh94} Robert, M.S. \& Haynes,
  M.P. 1994, \araa, 32, 115
\bibitem[Romanowsky et al.(2012)]{romanowsky12} Romanowsky, A.~J., Strader, J., Brodie, J.~P., et al.\ 2012, \apj, 748, 29
\bibitem[Rudick et al.(2011)]{rudick11} Rudick, C.~S., Mihos, J.~C., \& McBride, C.~K.\ 2011, \apj, 732, 48
\bibitem[Schaye et al.(2015)]{schaye15} Schaye, J., Crain, R.~A., Bower, R.~G., et al.\ 2015, \mnras, 446, 521
\bibitem[Strauss et al.(1992)]{strauss92} Strauss, M.~A., Huchra, J.~P., Davis, M., et al.\ 1992, \apjs, 83, 29
\bibitem[Silverman(1986)]{silverman86} Silverman, B., \ 1986, Density Estimation for Statistics and Data Analysis, (Chapman and Hall)
\bibitem[Springel et al.\ (2005)]{springel05} Springel, V., White, S.~D.~M., Jenkins, A., et al.\ 2005, \nat, 435, 629
\bibitem[Springel et al.(2008)]{springel08} Springel, V., Wang, J., Vogelsberger, M., et al.\ 2008, \mnras, 391, 1685
\bibitem[Tonry et al. (2001)]{tonry01} Tonry, J.L., Blakeslee, J.P.,
  Ajhar, E.A., Fletcher, A.B., Luppino, G.A., Metzger, M.R., \& Moore,
  C.B. 2001, \apj, 546, 681
  \bibitem[Tully (1988)]{tully88} Tully, R.B. 1988, Nearby Galaxies Catalog (Cambridge: Cambridge University Press) 
\bibitem[Tully et al. (2013)]{tully13} Tully, R.B., Courtois, H.M.,
  Dolphin, A.E., Fisher, J.R., Heraudeau, P., Jacobs, B.A.,
  Karachentsev, I.D., Makarov, D., Makarova, L., Mitronova, S., Rizzi,
  L., Shaya, E.J., Sorce, J.G., \& Wu, P.-F. 2013, \aj, 146, 86
\bibitem[van Zee et al. (2004)]{vanzee04} van Zee, L., Barton, E.J., \& Skillman, E.D. 2004, \aj, 128, 2797  
\bibitem[Vogelsberger et al.(2014)]{vogelsberger14} Vogelsberger, M., Genel, S., Springel, V., et al.\ 2014, \mnras, 444, 1518

\bibitem[Yahagi \& Bekki(2005)]{yahagi05} Yahagi, H., \& Bekki, K.\ 2005, \mnras, 364, L86
\bibitem[Young (2016)]{young16} Young, M.D., 2016, ProQuest Dissertations Publishing
\bibitem[Zepf \& Ashman(1993)]{za93} Zepf, S.E., \& Ashman, K.M. 1993,
  \mnras, 264, 611
\bibitem[Zinn (1985)]{zinn85} Zinn, R. 1985, \apj, 293, 424


\end{thebibliography}
\end{document}